\documentstyle[aps,multicol,epsfig]{revtex}
\newcommand{\nc}{\newcommand}
\nc{\be}{\begin{equation}}
\nc{\ee}{\end{equation}}
\nc{\bea}{\begin{eqnarray}}
\nc{\eea}{\end{eqnarray}}
\nc{\bean}{\begin{eqnarray*}}
\nc{\eean}{\end{eqnarray*}}
\nc{\mb}{\mbox}
\nc{\rnc}{\renewcommand}
\nc{\r}{\mb{\boldmath$r$}}
\nc{\x}{\mb{\boldmath$x$}}
\nc{\A}{\mb{\boldmath$A$}}
\nc{\sa}{\mb{\boldmath$a$}}
\nc{\sss}{\mb{\boldmath$\sigma$}}
\nc{\nab}{\nabla}
\nc{\X}{\sf x}

\renewcommand{\narrowtext}{\begin{multicols}{2} \global\columnwidth20.5pc}
\renewcommand{\widetext}{\end{multicols} \global\columnwidth42.5pc}

\begin{document}
\draft

\title{Nontrivial behavior of the Fermi arc in the staggered-flux ordered phase} 
\author{Koichi Hamada and Daijiro Yoshioka}

\address{Department of Basic Science, 
University of Tokyo, 3-8-1 Komaba, Tokyo 153-8902, Japan}
\date{October 16, 2003}
\maketitle

\begin{abstract} 
The doping and temperature dependences of the Fermi arc  
in the staggered-flux, or the $d$-density wave, ordered phase 
of the $t$-$J$ model are analyzed by the U(1) slave boson theory.  
Nontrivial behavior is revealed by the self-consistent calculation. 
At low doped and finite-temperature region,
both the length of the Fermi arc and the width of the Fermi pocket 
are proportional to $\delta$ 
and the area of the Fermi pocket is proportional to $\delta^2$. 
This behavior is completely different from that at the zero temperature, 
where the area of the Fermi pocket becomes $\pi^2 \delta$. 
This behavior should be observed 
by detailed experiments of  angle-resolved photoemission spectroscopy  
in the pseudogap phase of high-$T_{\rm c}$ cuprates 
if the pseudogap phase is the staggered-flux ordered phase.  
\end{abstract}



\begin{multicols}{2}



The essence of the pseudogap phase of the high-$T_c$ superconductors in the low-doping region has not been clarified.  
There are several theoretical proposals for the origin of this phase, 
\cite{WenLee96,CL01,Stripe99,AFflc96,SCphase97,SCpre94,Janko97,Yanase99,Rohe01,Allen01,Kyung01} 
and among them staggered flux state\cite{AM88,HMA91} characterized by staggered orbital current 
is one of the most promising candidate.  
This state successfully explains various aspects of the pseudogap phase, 
such as 
weak magnetic signals caught by recent neutron scattering experiments 
of underdoped YBa$_2$Cu$_3$O$_6$ (YBCO)\cite{Mook01,Mook0204}, 
alternating magnetic signals in the vortex core observed 
by recent muon spin rotation experiments of underdoped YBCO\cite{Miller}, 
structure of underdoped vortex, \cite{Lee_Wen01} 
lack of specific-heat anomaly,\cite{KeeKim02,Chakravarty0206} 
competition with $d$-wave pairing,\cite{UL92,koichi-physicac,koichi-prb} 
gap evolution,\cite{koichi-prb} 
etc. 
In particular, it has been shown that the Fermi arc observed by
the angle-resolved photoemission spectroscopy (ARPES) 
experiments\cite{arpes-review03} in the pseudogap phase can be explained
by the staggered flux state.\cite{koichi-prb,CNT03-ARPES} 
Here what is meant by the Fermi arc is a gapless region near ($\pi/2, \pi/2$), which does not form a closed loop. 

It is desirable, however, to have further support for the staggered-flux state.  
In a previous paper\cite{koichi-prb} we have noticed that the theoretically obtained Fermi arc, or Fermi surface, 
at finite temperature is much smaller than that of a naive expectation that it is given by the density of doped holes.  
This means that the Fermi arc has strong temperature dependence, 
and hence it should have nontrivial doping dependence at finite temperatures. 
Thus if such nontrivial behavior of the Fermi arc is observed experimentally, 
it will give additional support for the staggered flux state.  
In order to make the comparison between the theory and experiment meaningful, 
we have investigated the $t$-$J$ model, and obtained detailed temperature and doping dependence of the Fermi surface
in the staggered flux phase.

We have investigated the $t$-$J$ model based on the U(1) slave boson theory. 
In the present paper we set the parameter $t/J=3$ that is relevant for high-$T_{\rm c}$ superconductors
and the unit lattice length $a=1$. 
The exchange energy $J$ is conventionally considered to be around 1500K. 

We firstly review the proposed phases and the Fermi surfaces in the U(1) slave boson $t$-$J$ model. 
The uniform RVB (resonating valence bond) phase, where hopping order parameters are real and uniform,
is thought to represent the anomalous metal (non Fermi-liquid) phase  
above the Bose-condensation (BC) temperature of holons\cite{NL-uniformRVB}
and the Fermi-liquid phase below the BC temperature.   
In this phase, a large Fermi surface is formed whose area is proportional to 
$1- \delta$ and Luttinger theorem is satisfied, 
where $\delta$ is hole concentration. 
The $d$-wave RVB phase, where spinons form $d_{x^2-y^2}$-wave pairs, 
and $d$-wave superconducting ($d$-wave RVB phase with BC of holons) phase\cite{Kotliar88,Suzumura88}  
have always pointlike Fermi surface which exists in the nodal direction. 
In the staggered-flux ordered phase, the Fermi surface forms an arc\cite{koichi-prb}
consistent with the ARPES experiments\cite{arpes-review03} in the pseudogap phase. 
To be more precise, the Fermi arc is a natural feature only in the U(1) slave boson theory. 
On the other hand, in the SU(2) slave boson theory, artificial introduction of the phenomenological interactions are needed 
for reproducing a Fermi arc 
because the chemical potential is always zero due to the SU(2) symmetry even at finite doping.\cite{WenLee96}

In ordinary metals, the Fermi surface is almost temperature-independent because 
even room temperature is much lower than its Fermi energy.   
On the contrary, we reveal that there is strong and nontrivial doping and temperature dependences 
of the Fermi arc in the staggered flux ordered phase in this paper. 
%
This is because the low-energy excitation in this phase is described by 
a (2+1)-dimensional $anisotropic$ massless-Dirac-Fermion around the ($\pi/2, \pi/2$) (and its symmetric points) 
with finite chemical potential.
The expanded spectrum around the ($\pi/2, \pi/2$) is given as the following:  
\bea
E_{{\bf k} \pm} \approx& \pm \sqrt{2}\sqrt{ X^2  \tilde{k}_+^2 + Y^2 \tilde{k}_-^2 } -\mu, 
\eea  
where 
$X= 2t x_{\mathrm h} + (3J/4) x_{\mathrm s} $, 
$Y= 2t y_{\mathrm h} + (3J/4) y_{\mathrm s} $,
$\tilde{k}_{\pm}= (\tilde{k}_x \pm \tilde{k}_y)/ \sqrt{2}$, 
$\tilde{k}_{i}= k_{i} - \pi/2$ ($i = x, y$), 
and $x_{\rm s}$, $y_{\rm s}$, $x_{\rm h}$, and $y_{\rm s}$ are order parameters 
whose details will be elaborated later. 
Since $\mu < 0$, $ E_{{\bf k} -}=0$ gives the Fermi arc. 
As $X > Y$ in the staggered-flux phase,     
the zero-energy line of the fermion forms an ellipse at finite doping (Fig. 1).
The Fermi surface can be considered as an arc  
although the zero-energy line of the fermion forms an ellipse.   
It is because the intensity of the spectral function is small 
in the outer region (where $|k_{x}|+|k_{y}| \ge \pi $).\cite{koichi-prb}

\begin{figure}[t]
\begin{center}
\epsfxsize 8cm \epsffile{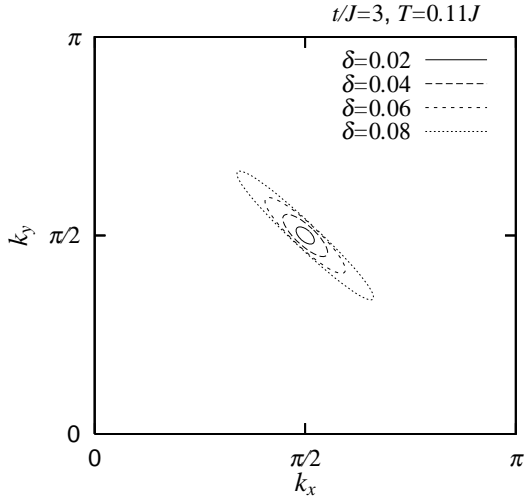}
\end{center}
\caption{
The doping dependence of the zero-energy line of the fermion at $T=0.11J$.
The first quadrant of the Brillouin zone is shown. 
The Fermi surface can be considered as an arc 
although the zero-energy line of the fermion forms an ellipse. (Ref. 22)}
\label{FS}
\end{figure}

First, we show doping-dependence of 
three characteristic quantities of the Fermi surface shown in Fig. 1:  
(i) the major axis $l_1$ in the $(0,\pi)$-$(\pi,0)$-direction, 
(ii) the minor axis $l_2$ in the $(0,0)$-$(\pi,\pi)$-direction, 
and (iii) the area of the Fermi pocket $S$. 

At zero temperature (Fig.2 (a)) the area of the Fermi pocket $S$ is $\pi^2 \delta$, 
which is determined by the self-consistent equation for total fermion number.  
The $l_1$ and $l_2$ are proportional to $\sqrt{\delta}$ near half-filling   
because the amplitude of the flux is near $\pi$, i.e., $X \approx Y$,  
but in other region the dependence is not $\sqrt{\delta}$. 
When the amplitude of the staggered flux decreases, 
the rate $Y/X$ decreases and the ellipse of the energy-contour 
develops to $\tilde{k}_x= - \tilde{k}_y$ ($(0,\pi)$-$(\pi,0)$) direction. 
When the staggered-flux order is completely destroyed, 
a large Fermi-surface whose area is proportional to $1-\delta$ is formed 
and $l_1$ becomes roughly $\sqrt{2} \pi$.

At finite temperature, the behavior is different.  
In Fig. 2 (b), we show a characteristic doping dependence at finite temperature, 
$T=0.11J$, which corresponds to about 165 K.  
Both the major axis $l_1$ and the minor axis $l_2$ develop linearly to $\delta$ 
and the area $S$ is proportional to $\delta^2$ at lightly doped region.   
As doping increases, $l_2$ saturates rapidly and only $l_1$ increases, 
i.e., the Fermi arc develops only in the $(0,\pi)$-$(\pi,0)$-direction.

\begin{figure} 
\begin{center}
\epsfxsize 9cm \epsffile{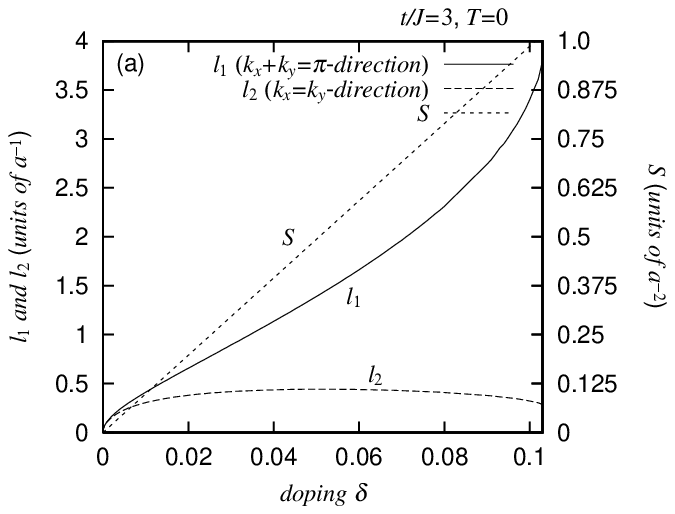}
\epsfxsize 9cm \epsffile{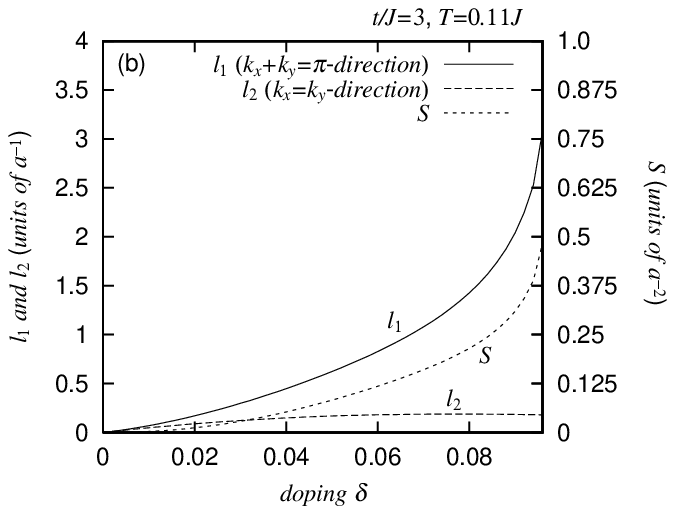}
\end{center}
\label{d-dep}
\caption{ 
A characteristic doping-dependence of the Fermi arc at low and high temperature region:  
The dependences of the major axis $l_1$, the minor axis $l_2$, 
and the area $S$ at $T=0$ (a) and $0.11J$ (b) are shown.
}
\end{figure}

This theoretically obtained Fermi arc, or Fermi surface, at finite temperature (Fig. 2(b)) 
is much smaller than that of a naive expectation that it is given by the density of doped holes.  
This means that the Fermi arc has strong temperature dependence. 
The temperature dependence at $\delta=0.01$ is shown in Fig. 3. 
When temperature increases, the area $S$ becomes smaller.

In the staggered flux phase, the low-energy-excitation is 
described by an $anisotropic$ massless-Dirac-Fermion with finite chemical potential 
and the shape of the spectrum is conical. 
As the density of states of the lower cone, where the Fermi level lies, decreases rapidly at higher energy, 
the chemical potential increases as the temperature becomes higher, 
and the cross section of the cone becomes smaller. 
This is the origin of the strong temperature dependence of the Fermi arc at low temperature.
The length of the Fermi arc stops to decrease and begins to increase when temperature increases much higher,  
as the staggered flux order begins to be destroyed 
with approaching the transition temperature to the uniform RVB phase. 
Finally, at the transition temperature, 
a large Fermi-surface is formed.  

When the doping increases, this temperature dependence of the Fermi arc wholly becomes weaker.  
It is because 
the amplitude of the staggered flux order decreases and the conical spectrum becomes flatter.
As the minimum value of $l_1$ becomes larger and $l_1$ is always fixed roughly $\sqrt{2} \pi$ at transition temperature, 
the sharpness of the increase near the transition point also becomes softer.     
These strong temperature dependences also make the doping dependence of the Fermi arc nontrivial.
\begin{figure}
\begin{center}
\epsfxsize 9cm \epsffile{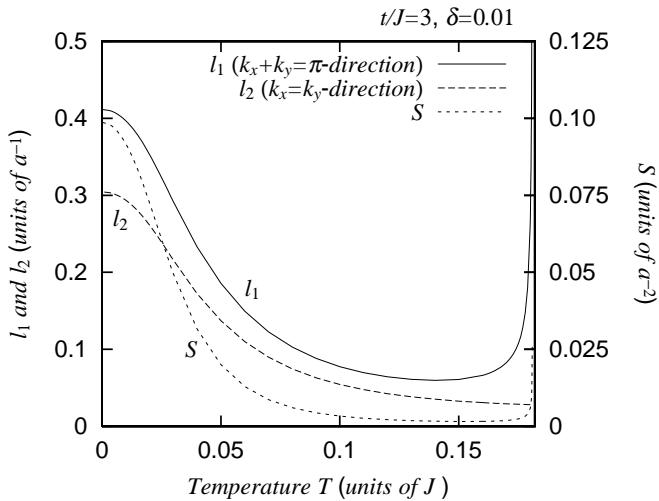}
\end{center}
\caption{ 
A characteristic temperature-dependence of the Fermi arc:  
The dependence of the major axis $l_1$, the minor axis $l_2$, and the area $S$ at $\delta=0.01$ are shown.
}
\label{T-dep}
\end{figure}

In the following we report the details of our calculation. 
We follow the same path taken in Ref. \ref{koichi-prb} while 
considering only the staggered-flux state here, 
not considering the $d$-wave RVB or the $d$-wave superconducting state. 

We analyze the two-dimensional $t$-$J$ model on a square lattice: 
$H = \ -t \sum_{ \langle i,j \rangle } \sum_{\sigma} 
\mbox{\boldmath $P$}(c_{i\sigma}^{\dagger}c_{j\sigma}+{\rm H.c.}) \mbox{\boldmath $P$}
+ J \sum_{\langle i,j \rangle} \mbox{\boldmath $S$}_{i}\cdot\mbox{\boldmath $S$}_{j}$
that describes doped Mott insulators which is essential of high-$T_c$ superconductors. 
Here, $\langle i,j \rangle$ represents sum over the  nearest-neighbor sites  
and {\mbox {\boldmath $P$} } is a projection operator to no doubly occupied state,  
$\mbox{\boldmath $S$}_i $ represents spin-1/2 operator 
$\mbox{\boldmath $S$}_i=\frac{1}{2}c_{i\sigma}^{\dagger}(\mbox{\boldmath $\sigma$})_{\sigma \sigma^{'}}c_{i\sigma^{'}}$,  
where \mbox{\boldmath $\sigma$}$=(\sigma_1, \sigma_2, \sigma_3)$ are Pauli matrices.
In the U(1) slave boson theory, 
the physical electron operator $c_{i\sigma}$ is described by a product of    
an auxiliary spin-1/2 neutral fermion operator $f_{i\sigma}$ called spinon  
and an auxiliary spinless charged boson operator $b_{i}$ called holon;  
$c_{i \sigma}=b^{\dag}_{i}f_{i\sigma}$ with a constraint  
$b_{i}^{\dagger}b_{i}+ \sum_{\sigma} f_{i\sigma}^{\dagger}f_{i\sigma}=1$.

For the mean-field solution, 
we considered both the staggered flux order of spinons and that of holons: 
$ \langle f_{ i+ {\hat x} \sigma}^{\dag}f_{i\sigma} \rangle  
= \chi {\mathrm e}^{i(-1)^{i} \phi_{\mathrm s}/4 }
 = x_{\mathrm{s}} + i(-1)^{i} y_{\mathrm{s}}, 
\langle f_{ i+ {\hat y} \sigma}^{\dag}f_{i\sigma} \rangle  
=\chi {\mathrm e}^{-  i(-1)^{i} \phi_{\mathrm s}/4 } 
= x_{\mathrm{s}} - i(-1)^{i} y_{\mathrm{s}},  
\langle b^{\dag}_{i+ {\hat x}} b_{i} \rangle 
= B {\mathrm e}^{i(-1)^{i} \phi_{\mathrm h}/4 } 
= x_{\mathrm{h}} + i(-1)^{i} y_{\mathrm{h}},$ 
and 
$\langle b^{\dag}_{i+ {\hat y}} b_{i} \rangle 
= B {\mathrm e}^{-i(-1)^{i} \phi_{\mathrm h}/4 } 
= x_{\mathrm{h}} - i(-1)^{i} y_{\mathrm{h}} $. 
Here, ${\hat x}$ and ${\hat y}$ are unit vectors in the $x$ and $y$ directions, 
$x_{\mathrm{s}}= \chi \cos(\phi_{\mathrm{s}}/4)$, 
$y_{\mathrm{s}}= \ \chi \sin(\phi_{\mathrm{s}}/4)$,  
$x_{\mathrm{h}}= B \cos(\phi_{\mathrm{h}}/4)$, and 
$y_{\mathrm{h}}= B \sin(\phi_{\mathrm{h}}/4)$.  
The staggered flux state contains a density wave (particle-hole pairing) ordering
whose symmetry is $d_{x^2-y^2}$, which is called ``$d$-density wave.''
\cite{Nayak00} 
The order parameters $y_{\mathrm{s}}$ and $y_{\mathrm{h}}$ correspond to the  
$d$-density-wave order parameter of spinons and holons, respectively. 
The mean-field Hamiltonian is diagonalized by the unitary transformation of the spinon and the holon. 
The obtained spectrum of fermions 
is $ E_{{\bf k} \pm}= \pm \sqrt{ \epsilon_{{\bf k}}^{2} + W_{{\bf k}}^{2} } -\mu^{\mathrm}$, 
where $\epsilon_{{\bf k}} =- X (\cos k_{x}+ \cos k_{y})$ and 
$W_{{\bf k}}= Y (\cos k_{x}- \cos k_{y})$. 
%
We solved self-consistent equations numerically and the results are shown in the figures.  
The line of the $ E_{{\bf k} -} =0$  
is shown in Fig. \ref{FS}.  
The upper band spectrum $E_{{\bf k} +}$ is always positive at finite doping where $\mu < 0$.

%
In the present paper, we neglected $d$-wave RVB order  
because it was shown that the $d$-wave RVB order is completely destroyed 
by the U(1) gauge fluctuation 
above the BC temperature of the holons.\cite{UL94} 
This means that the pseudogap region of the $t$-$J$ model, 
which is the underdoped region above the BC temperature, 
cannot be explained by the $d$-wave RVB state.  
We think that only possible state in this region is 
the staggered-flux state, 
which is equivalent to the $d$-wave RVB state at half-filling due to a SU(2) symmetry.\cite{Affleck88} 
The reasons were discussed elsewhere.\cite{koichi-prb} 
It was shown by a variational Monte Carlo method that the ground state
of the $t$-$J$ model near half filling is antiferromagnetic (AF) state.\cite{Yokoyama96}
However, in this paper we also neglected the possibilities of the AF state  
because it is known that experimentally there is no AF spin order in the pseudogap phase that 
we focused in this paper. 
We adopt the $t$-$J$ model where only the nearest-neighbor hopping term, often called the $t$ term, 
is included in the present paper.  
For a better description of actual high-$T_{c}$ superconductors, 
it is sometimes claimed that the next-nearest-neighbor hopping term, the $t^{'}$ term, is necessary.\cite{Tanamoto}
We think that the $t^{'}$ term does not change the results qualitatively 
as long as the shape of the spectrum remains conical around $(\pi/2, \pi/2)$.   
In the staggered flux phase, it should be so. 
%
The growth of the Fermi arc near the transition point is sharp because 
our analysis is based on a mean-field theory.   
However, it is expected that the sharpness becomes weaker 
when the fluctuation around the mean-field solution is included.
%

In conclusion, we have analyzed 
doping and temperature dependences of 
the Fermi arc in the staggered flux ordered phase. 
Nontrivial behavior has been revealed. 
This behavior should be observed 
by precise experiments of ARPES 
in the pseudogap phase of high-$T_{\rm c}$ cuprates, 
if the pseudogap phase is the staggered-flux ordered phase.  

K.H. thanks Masao Ogata, Youichi Yanase, Yoshifumi Morita, Takashi Koretsune, and Naokazu Shibata  
for their useful discussions. 
D.Y. appreciates hospitality of Aspen Center for Physics, where part of this work was done. 
Numerical computation in this work was partially carried out 
at the Yukawa Institute Computer Facility.


\end{multicols}

\end{document}